# Electron Transport in Hybrid Metallic Nanostructures
## (Metallic Nanoelectronics)

V. T. Petrashov

*Department of Physics, Royal Holloway, University of London, Egham, Surrey TW20 0EX, United Kingdom*

A summary is given of a talk [1] on the physics and technology of hybrid metallic nanostructures, with a view to metallic nanoelectronics. In the beginning of the talk it was noted that the majority of the presentations at the conference were not concerned with electronics but the physics of sensors, since they were devoted to studies of the influence of temperature, magnetic field, *e.g.* [2], whereas electronics is all about modulation of electrical conductance by electrical means. An overview was given of the latest research into electric field effects in metals, including quantum field effects [3,4] and effects in ultra-thin quench-condensed metallic films [5]. A design of Metallic Field Effect Transistor using 2D metallic films was presented, and an estimate for electric field effect was given. It was emphasized that 2D metallic nanostructures may show completely different properties than bulk metals and are in essence novel materials. Examples of metallic systems that could show field effects in accessible electric fields were given. A nanotechnology [6] enabling fabrication of thermodynamically stable ultra-thin 2D metallic nanostructures was presented; a MFET of similar design has been implemented in [7] using graphene as 2D metal.

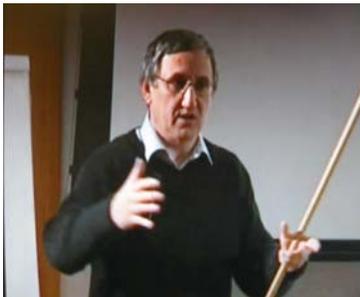

*1. Why metals?* The title of this conference is "nanoelectronics" so it is a good time to consider seriously one of the scenarios *for nanoelectronics*, namely *the metallic scenario.* Electronics is all about *modulation of the conductance by electrical means*; the control using magnetic field, temperature *etc.* presented in many talks here is not about nanoelectronics, but sensors. In this talk I will summarise what has already been achieved by *electric field control* and the effects that can be observed in metals within current technology. I will also discuss the *technological challenges of metallic nanoelectronic devices.* The prime question is *"why metals?"* If you consider devices made of semiconductors with *all three dimensions* at the nanometre scale you find them in the *insulating regime* because there are so few electrons. We run into a crisis similar to that with miniature vacuum electronic devices. There are too few electrons, about $10^{10}$ cm$^{-3}$ can be generated in vacuum, and you cannot make a reliably operating device on the millimetre scale. It was the ability to control the conductance in semiconductors with much higher concentrations of $n \sim 10^{18}$ cm$^{-3}$ electrons that led to a revolution in electronics. So if we follow "the smaller the device the higher concentration of electrons you need" trend, you have an answer to "why metals*?": in metals the concentration of electrons is up to five orders of magnitude higher than in semiconductors.* More than 80% of the Periodic Table are metals and all relevant microelectronics length scales in metals are at the nanometre scale. This includes the mean free path of electrons at room temperature, the optical skin depth, and the screening radius.

*2. Quantum metallic nanostructures.* There have already been two generations of metallic nanostructures. The first generation was just plain single layer metallic nanostructures that showed quantum interference effects [8-11]. When it was possible to align different metallic layers with high precision during lithography much more complicated hybrid metallic nanostructures of the second generation were fabricated showing interesting new physics [12].

So what about electric field effects? The *h/e quantum oscillations* in metallic nano-rings [9] were found to be sensitive to electric field via the *electric field Aharonov-Bohm effect*, and were investigated in Antimony rings by the group of Webb in eighties [3]. Another experiment was made with nano-rings made of Bismuth [4]. The rings were similar to those used in Ref. 3, with extra electrodes to squeeze conducting channels using needle-like gates (Fig. 1). The Universal Conductance Fluctuations in the applied electric field were observed (Fig. 1). It was also established that screening was strongly suppressed in small metallic wires.

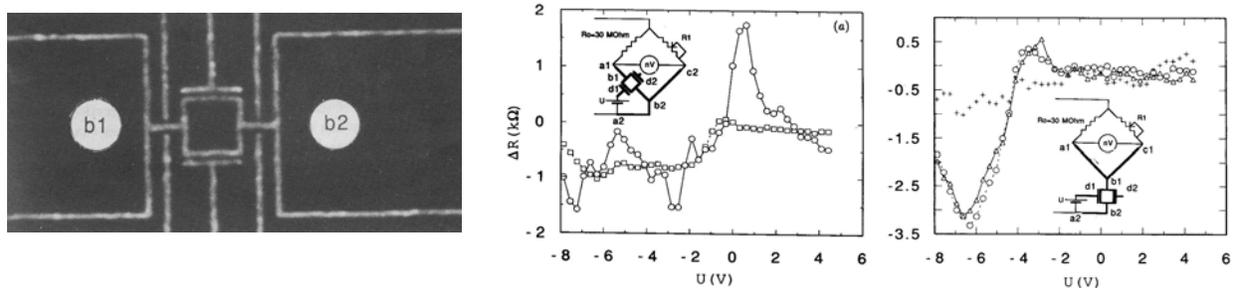

Fig. 1 Field effect in mesoscopic Bi wires at different temperatures and sample configurations [4].

The quantum electric field effects can be explained by a simple picture. If you pump electrons into the sample you change the Fermi wavevector, and hence the phase of conduction electrons. This may change the interference from constructive to destructive in a relatively low electric field, provided the electron coherence length is long enough.

One of the examples of hybrid metallic nanostructures is the *Andreev interferometer* with cross-like normal metallic part A-B-C-D with interfaces to a superconducting wire (Fig. 2) at points D and C. The electrical conductance between points A and B can be manipulated by tuning the superconducting phase difference between points D and C using control current in the superconducting wire (Fig. 2) [13]. This is an essentially low temperature electronic device where the conductance is controlled by an electrical current.

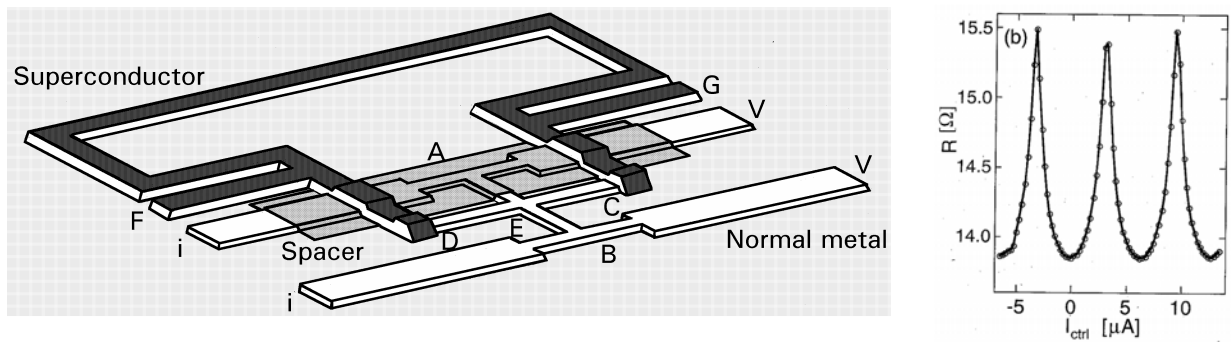

Fig 2. Control of electrical conductance in a metallic mesoscopic conductor A-B by electrical current in adjacent superconducting wire F-G [13].

*3. Electric field effect in 2D metals.* Next I would like to re-visit an idea of metallic nano-electronics with "classical" electric field control that can survive up to room temperature.

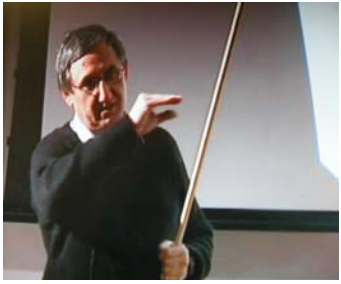 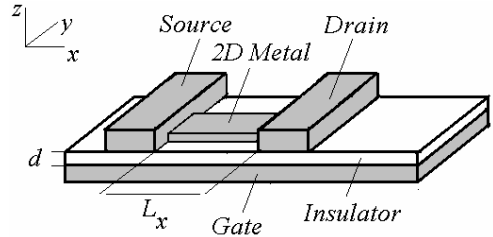

Fig. 3. Metallic Field Effect Transistor

A device is shown in Fig. 3, and consists of a thin metallic bridge of thickness $L_z$, a gate, and an insulator of thickness $d$. The extra charge that can be induced in the bridge is *limited by the breakdown electric field, $E_0$*:

$$\delta Q_{max} = e\delta N_{max} = CV_0 = \varepsilon \frac{L_x L_y}{d} V_0 = \varepsilon A E_0 \qquad \left\langle \frac{\delta n}{n} \right\rangle_{max} = \frac{\varepsilon E_0}{e L_z n}.$$

From this one can estimate the change in the Fermi energy averaged over the thickness of the bridge:

$$\left\langle \frac{\delta \varepsilon_F}{\varepsilon_F} \right\rangle_{max} = \frac{2}{3} \frac{\varepsilon E_0}{e L_z n} \quad \text{for} \quad n = \frac{1}{3\pi^2} \left( \frac{2m\varepsilon_F}{\hbar^2} \right)^{3/2}.$$

To achieve an electric field effect with realistic breakdown electric field one needs a thin single atomic layer metallic bridge. If you use some exotic insulator you can pump in more electrons; however they will be accumulated in a *thin layer of the order of screening depth.* In a thick metallic film this layer will be short circuited by the bulk electrons**.** So the film has to be *several atomic layers thick.*

Uniform one-single-atomic-layer-thick quench-condensed metallic films have already been fabricated [5] and field effects have been observed (Fig. 4). However such films were not really metallic films, they had sheet resistances of order of 40 kΩ and exhibited a complicated glass-like behaviour.

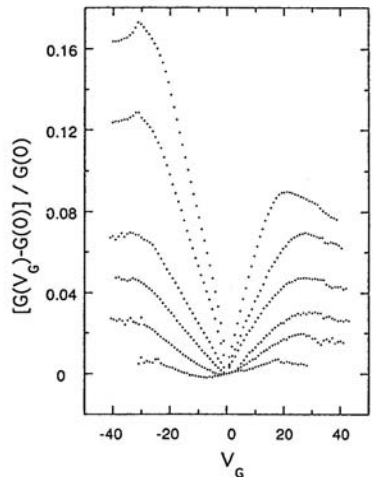

Fig. 4. Electric field effect in ultra-thin quench-condensed metallic films [5].

I am going to talk about ordered ultra-thin metallic films.  For a single atomic metallic layer the sheet resistance is expected to be 10 - 50 Ω. It is a quite difficult technological problem to make such films. There are some metallic candidates for observation of the field effect even without the essential modulation of the concentration.

One example is Nickel. In Fig. 5 you see the results of calculations for the density of states in Nickel for the majority and minority spins. There are different densities for spin up and spin down states near the Fermi level leading to spin polarisation, and a slight shift of the Fermi level will drastically change the spin polarisation. This should lead to interesting field effects in electron tunnelling and electron transport through hybrid ferromagnetic/superconducting nanostructures. This is just one simple case of about a hundred elements in periodic table.

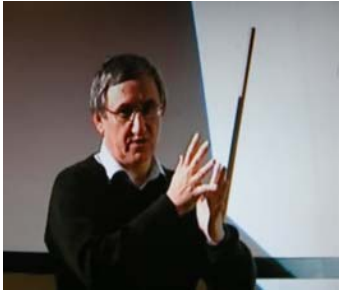
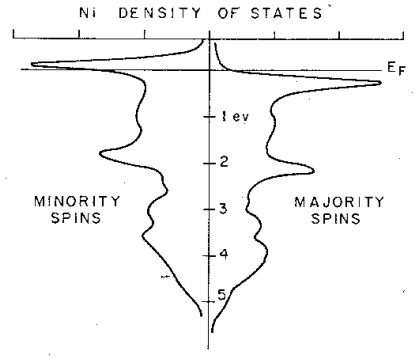

Fig. 5. Electron density of states in Nickel.

Example 2. This involves complicated materials, f-electron metals, which are very sensitive to the pressure, for example Lanthanum with Cerium. It can go from magnetic order to superconducting state (Fig. 6). Such systems would be good candidates for electric field effect since the change in Fermi wave-vector by electric field effect and the change in the reciprocal lattice vector due to application of pressure may lead to similar results. Many other metallic systems can be explored.

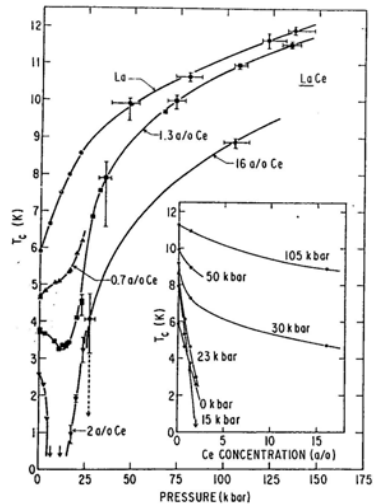

Fig. 6. Pressure-sensitive transport properties in f-electron materials. Superconducting critical temperature $T_c$ *vs* pressure in the $La_{1-x}Ce_x$ system [13].

*4. Nanotechnology for ultra-thin metallic films.* For stable ultra-thin metallic systems we suggest a technology based on *direct write with a collimated atomic nano-beam* [6]. We use a membrane or a double layer resist system with a pin-hole as a mask (Fig. 7 a). Placing the mask attached to a substrate onto a computer-controlled holder, which can be rotated in pre-programmed way, we can in principle write any structure. One more process in this technology is self-narrowing of beams that go through the pin-hole due to the build-up of material (Fig. 7 b). You can use this process to make much smaller features than pin-holes originally made by e-beam. Fig. 7 c shows very thin Silver ring a couple of nanometres thick. Since thermal radiation is reflected by the membrane and the atomic beam is very narrow, the surface of the substrate is cold enough to suppress diffusion. As a result the ring is continuous, even at a small thickness. Furthermore, we were able to place aluminium islands onto the Silver ring with high precision by switching evaporation sources and "drawing" different parts of the device without breaking the vacuum. The technique allows the fabrication of many other kinds of nanostructures that are not attainable with traditional technologies. Fig. 7d. shows a *structure with a gate and ultra-thin metallic bridges and electrodes for measurements of electric field effect.* We are going to measure this kind of devices and I hope we are able to report results soon [14].

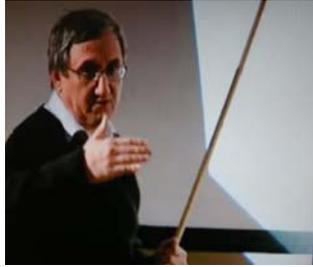

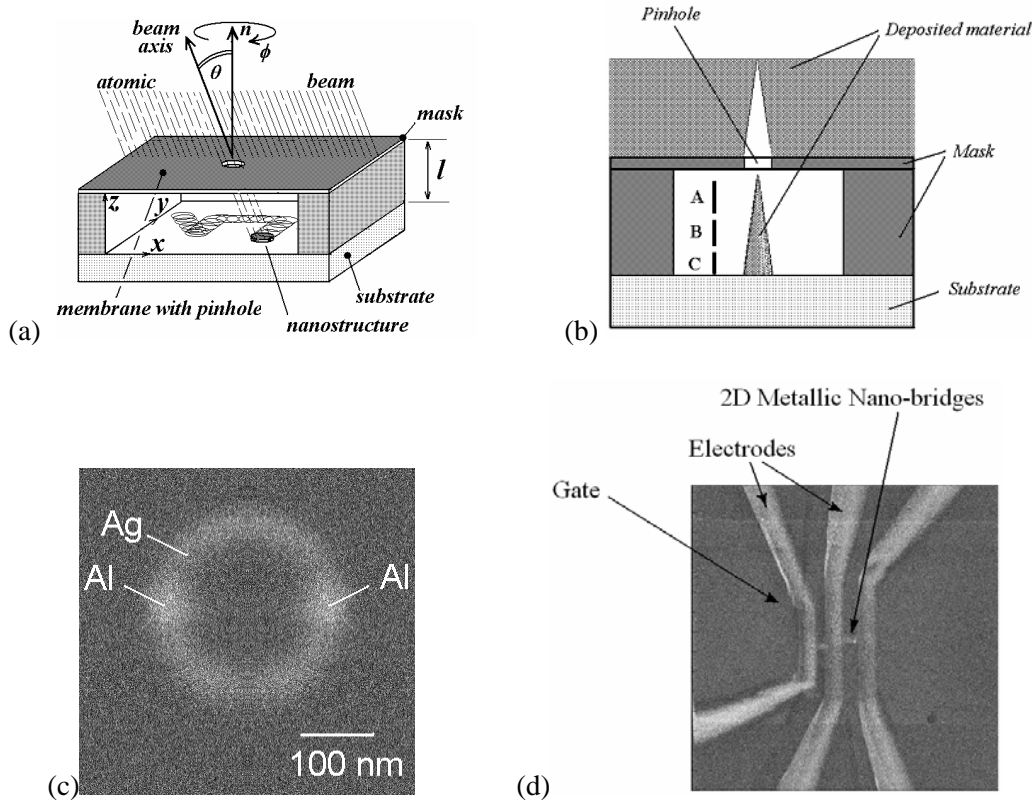

Fig. 7. Nanofabrication of stable ultra-thin metallic structures using collimated atomic beams [7, 11]

*5. Conclusions.*

- *2D Hybrid Metallic Nanostructures have a great potential for both, fundamental research and future applications*
- *2D Metallic Nanostructures may show completely different properties, and are in essence novel materials even fabricated using elementary metals.*
- *Electric Field Effects may open new possibilities in control of electron transport and magnetic properties of 2D ferromagnetic nanostructures.*
- *Further development of novel technologies for fabrication of homogeneous, highly ordered metallic 2D nanostructures is in order.*

*Acknowledgements.* I thank the organizers, especially C. Lambert and V. Falko, for inviting me to give a talk at this momentous conference.